\documentclass[8pt,aps,twocolumn,showpacs]{revtex4}
\usepackage{amssymb}
\usepackage{latexsym}
\usepackage{epstopdf}
\usepackage{amsmath}
\usepackage[mathcal]{eucal}
\usepackage{graphicx}
\usepackage{dcolumn}
\usepackage{amsmath}
\usepackage{amsfonts}
\usepackage{bm}

\newcommand{\be}{\begin{equation}}
\newcommand{\ee}{\end{equation}}
\newcommand{\bea}{\begin{eqnarray}}
\newcommand{\eea}{\end{eqnarray}}

\newcommand{\p}{\partial}
\newcommand{\s}{\sigma}

\newcommand{\la}{\langle}
\newcommand{\ra}{\rangle}
\newcommand{\rd}{\mbox{d}}
\newcommand{\ri}{\mbox{i}}
\newcommand{\re}{\mbox{e}}

\begin{document}

\title{Influence of thermal phase fluctuations on the spectral function in a 2D d-wave superconductor}

\author{M. Khodas and A.M. Tsvelik}
\affiliation{Department of Condensed Matter Physics and Materials Science, Brookhaven National Laboratory,
  Upton, NY 11973-5000, USA}

\begin{abstract}
We study the spectral function of  a two-dimensional superconductor in a regime of strong phase fluctuations. 
Although the developed approach is valid for any symmetry, we concentrate  on $d$-wave superconductors. We obtain analytical expressions for the single electron Green function below the transition temperature and have worked out a way to extrapolate it  for a finite temperatures above $T_c$. The suggested approach are easely generalizable for other models with critical fluctuations. 
  \end{abstract}


\maketitle
\section{Introduction}

 According to a popular viewpoint the cuprate superconductors in their underdoped regime are fundamentally different from  the BCS ones due to the  existence of wide temperature range  around  $T_c$ dominated by phase fluctuations. In this range the order parameter amplitude is well defined but the global order has not yet emerged. The quantitative measure of this difference is the ratio $Q= 2T_c/\pi\rho_s(0)$ where $\rho_s(0)$ is  zero temperature  phase stiffness. This ratio determines how close the transition is to being  mean-field-like. In BCS superconductors $Q\ll 1$, in the underdoped cuprates $Q \sim 1$ \cite{EK}. Experimental evidence in favor of strong phase fluctuations comes  from measurements of diamagnetic susceptibility and Nernst effect above $T_c$\cite{Ong} and  from temperature dependence of the thermal expansion coefficient \cite{exp}. The analysis of temperature dependence of magnetization and London penetration depth for the high quality underdoped  BiSCO crystals show that the superconducting  transition itself becomes closer to two-dimensional Berezinskii-Kosterlitz-Thouless (BKT) one \cite{Cam}. At last, there is a material (x=1/8 LSCO) where the coupling between the CuO planes is so weak that the transition is real BKT \cite{Tranq}.

 The problem of influence of the  phase fluctuations on the spectral function is a long standing one; its importance being  substanciated by the fact that the spectral function measured  by Angle Resolved Photoemission (ARPES) serves as one of the main probes of the strong correlations in the cuprates. The experiments show that in the underdoped regime many  features of the spectral function below $T_c$ survive above $T_c$ though in somewhat modified form.  According to ARPES, the excitation spectrum above $T_c$ gradually softens and loses its characteristic node-centered conical shape so that the nodal points gradually broaden into  arcs \cite{ARPES},\cite{ARPES1},\cite{Gap}. The question is whether the appearance of these arcs (or even of the  Fermi pockets) is due to superconducting fluctuations, as suggested, for instance, in \cite{PWA}. 
  To have a consistent understanding  of the underdoped regime  it is vital to obtain a correct detailed picture of the quasiparticle spectral function and establish its temperature dependence. The first step in this direction is to take into account superconducting fluctuations. This is possible to do even without full microscopic theory. The problem has been studied by a number of authors (\cite{Millis},\cite{Dorsey},\cite{Berg}), but, as we argue in this paper, the approach taken  is unreliable being based on uncontrolled approximations.      

\section{Formulation of the model}

 We consider a two-dimensional metal with a strong superconducting pairing  in the state where the order parameter amplitude is fixed, but the phase fluctuations are strong. Our calculations allow for the  SC (superconducting) gap to have  nodes on the Fermi surface.  For instance, for $d$-wave SC  the mean field quasiparticle spectrum is given by 
\bea
E^2 = \epsilon^2(k) + \Delta^2[\cos (a k_x) - \cos (a k_y)]^2 
\eea
We will assume that  $v/a \gg \Delta$ ($v$ is the Fermi velocity) and approximate the spectrum close to the node as 
\bea
E^2 \approx v^2k^2 + 2\Delta^2\sin^2(qa/2) \label{disp}
\eea
 where $k$ is the wave vector component perpendicular to the Fermi surface and $q$ is the one parallel to it. We take  the  Fermi surface at the node for a flat one.
 
In the mean field approximation fluctuations of the order parameter $\Delta$  are ignored, and the resulting Green function  takes a familiar BSC form,
$G_{k}(\omega) = (\omega + \epsilon_k )/( \omega^{2} - \epsilon_k^{2} - \Delta^{2}(k) )$. Following our original assumption we will  neglect the amplitude fluctuations of the order parameter taking into account only phase fluctuations,
$\Delta({\bf x},t) = \Delta \re^{ \ri \phi({\bf x},t) }$.  At finite temperature $T$ the  long wavelength  fluctuations are essentially classical (time independent). It is crucial for our consideration that the superconducting fluctuations are space isotropic and the one-particle Green function close to the node is strongly anisotropic with the parameter of anisotropy $\Delta/\epsilon_F$, where $\Delta$ is maximal gap and $\epsilon_F$ is the Fermi energy.  Then when one considers a propagating quasiparticle,   the fluctuations affect mostly its wave vector component perpendicular to the Fermi surface.  The parallel component can be considered as conserved. 
The above considerations allow us to consider one dimensional fermions at a given Matsubara frequency $ \omega_{n}$ 
described by the Lagrangian
\begin{align}
\mathcal{L} = 
\bar{\chi}_{\omega} 
\left[ - \ri \omega_{n} \hat{1}_{2} \! -\! \ri v \partial_{x} \sigma^{z} \!+\! \Delta \sigma^{+} \!+ \!\Delta^{*} \sigma^{-}   \right] \chi_{\omega}  
%
+
F_{\phi} \, . \label{action}
\end{align}
In the last equation the first term is a standard pairing Lagrangian written in terms of Nambu spinors,
$\chi_{\omega} = (\psi_{\uparrow,\omega_{n}}, \bar{\psi}_{\downarrow,-\omega_{n}} )^{T}$, and $\sigma^{i}$ are Pauli matrices.
The second term gives  the action for the classical phase fluctuations in the form
\begin{align}\label{2d}
\frac{F_{\phi}}{T} = \frac{\rho_s}{2T}\int dx dy \left[ (\p_{x}\phi)^2 + (\p_{y}\phi)^2  \right] \, . 
\end{align}
Here the inverse temperature prefactor $T^{-1}$ results from the integration over imaginary time.
The discrete symmetry of the lattice which includes $C_{4}$, the group of in-plane rotations by $\pi/2$ radians 
guarantees that the quadratic action, Eq.~\eqref{2d} is isotropic.

Remarkably, it is possible to rewrite the model, Eqs.~\eqref{action}, \eqref{2d}  
as an effective quantum impurity model of Caldeira-Leggett type \cite{cl}.
To this end, we introduce new variables as follows,
$[ \psi_{\uparrow}, \psi^{\dagger}_{\downarrow} ] = [ \Psi_{\uparrow}, - \ri \Psi^{\dagger}_{\downarrow} ] $,
and
$[ \psi^{\dagger}_{\uparrow}, \psi_{\downarrow} ] = [ \ri \Psi_{\uparrow},  \Psi^{\dagger}_{\downarrow} ] $.
Then, the Lagrangian Eq.~\eqref{action}, \eqref{2d} 
after the analytic continuation, $\ri \omega_{n} \rightarrow \omega + \ri 0$
is equivalent to the Hamiltonian
\begin{align}
H_{eff} = & v^{-1}\ri(\omega +\ri 0)\tau^3 + v^{-1}\Delta(q)[\tau^+\re^{\ri\phi(0)} + \tau^-\re^{-\ri\phi(0)}] \nonumber\\
& + H_{bulk}[\phi]\, , \label{Leggett}
\end{align}
where $\tau^a$ is the short hand notation for the fermionic bilinears:$
\tau^a \equiv \Psi^+\sigma^a\Psi$. 
In this setting coordinate $x$ plays the role of Matsubara time. It is dual to the momentum component $k_{\parallel}$ parallel to the Fermi velocity at the point of observation. Since at $\Delta/\epsilon_F \rightarrow 0$ the electron momentum parallel to the Fermi surface is conserved, the  fermionic field $\Psi_{\uparrow}, \Psi^{\dagger}_{\downarrow}$ depends only on one coordinate $x$, though the phase field $\phi$ depends on both. For convenience we assign $\Psi$ to coordinate $y=0$. Since the propagators of $\phi$-fields are not supposed to be modified by insersions of fermionic loops, which would lead to overcounting, the fermionic number must be treated as conserved $\psi^+_{\s}\psi_{\s} =1$. Then the $\tau$-operators become components of spin S=1/2. The Hamiltonian $H_{bulk}$ describes the phase fluctuations. Their two point correlation function is 
\bea
\la \re^{\ri \phi({\bf x}_1)}\re^{-\ri\phi({\bf x}_2)}\ra = \left| \frac{a}{\xi(T)}\right|^{2d}F\left(\frac{{\bf x}_{12}}{\xi(T)}\right) \, , \label{2point}
\eea
where $d = T/8T_{\mathrm{BKT}}$ is the order parameter scaling dimension, $\xi(T)$ is the correlation length and 
$T_{\mathrm{BKT}} = \pi \rho_{s}/2$ is the Berezinskii-Kosterlitz-Thouless transition temperature. The function $F(y)$ is such that $F(y \ll1) =y^{-2d}$ and $F(y > 1) \sim K_0(y)$ (more detailed information about this function can be extracted from \cite{LukZam}). Hence below the transition where $\xi = \infty$ the function (\ref{2point}) decays as a power law and above the transition it has the exponential asymptotics. The length scale $a \sim (v/\epsilon_{F})$ is the short distance cut-off.  Below the BKT transition the bulk Hamiltonian is Gaussian:
\bea
 H_{bulk}[\phi;\!T\! < \!T_{\mathrm{BKT}}]\! =\! \frac{1}{8\pi d}\int_{-\infty}^{\infty} \rd y 
 \left[  (4 \pi d)^{2}  \Pi^2 + (\p_y\phi)^2 \right], \label{Gauss}
\eea
where $\Pi$ is the momentum density conjugated to the field $\phi$, with an equal time commutator
\bea
\left[ \Pi(y_{1}),\phi(y_{2}) \right]_{-} = - \ri \delta(y_{1} - y_{2}) \, .
 \label{conjugate}
\eea
By construction integration over the momentum in the quantum action of our one-dimensional model, Eqs.~\eqref{Gauss}, \eqref{conjugate}
produces the free energy of classical two-dimensional thermal fluctuations, Eq.~\eqref{2d}.
Above the transition the Hamiltonian Eq.~\eqref{Gauss} must include effects of vortices which generate nonlinear terms. 
The rigorous discussion is possible in the case $T < T_{\mathrm{BKT}}$; later we will present some extrapolation for temperatures  above $T_{\mathrm{BKT}}$. 
We note in passing that at $T < T_{\mathrm{BKT}}$ model (\ref{Leggett},\ref{Gauss}) is equivalent to the anisotropic Kondo model in the imaginary magnetic field $h = \ri\omega - 0$. We notice this equivalency though we have not been able to make use of it in our calculations. 

The setting of the problem as given by Eqs.~\eqref{action}, \eqref{2d} is not different from the one in \cite{Millis},\cite{Dorsey},\cite{Berg}. However, in these previous attempts to solve the problem the authors used the gauge transformation method which we claim is inadequate. The bosonic exponent present in Eqs.~\eqref{action}, \eqref{Leggett} has been absorbed into the definition of the fermionic field. As a result the term 
$\psi^+_{\s}\psi_{\s}\p_x\phi(0)$
appeared in the Hamiltonian. The problem with this approach is that the path integral expression for the electron Green's function is dominated by the field configurations with large $\phi$ gradients. The same effect appears when one attempts to develop a perturbation theory  in $\p_x\phi$:  each diagram diverges at small distances. This fact has been overlooked.  Here the equivalence of the current problem to the Kondo model is helpful since  as is well known the latter cannot be treated by the methods employed in \cite{Dorsey},\cite{Berg}.

\section{Solution by perturbation theory in $\Delta$}

In the present section we calculate the Green function by 
the perturbative expansion in $\Delta$ for the model Eqs.~\eqref{action}, \eqref{2d}. 
It has a crucial advantage of being free of ultraviolet divergencies.
The infrared divergencies are removed at finite external frequency and momentum 
(as we will show below, the integrals diverge only at $\omega +vk =0$). 
The infrared behavior is controlled by the long wavelength fluctuations of the order parameter which justifies the  use of the effective low energy action for the phase fluctuations Eq.~\eqref{2d}. 
The above arguments indicate that we can study the model Eqs.~\eqref{action}, \eqref{2d} 
in perturbation theory expanding the Green function in powers of $\Delta$ by accounting for most infrared singular contributions at each order. 

Below for definiteness we consider positive energies, $\omega>0$.  
In other words we study the ``particle'' part of the spectrum. 
For negative frequencies the spectral function can be obtained by particle-hole transformation, $\omega \rightarrow - \omega$,
and $k \rightarrow - k$.
Most of the spectral weight is expected to be found close to the particle mass shell,
$k = \omega$.
In general, the Green function $ G^{r}_{\omega}(k) $ 
is peaked for wave vectors  close to the particle (hole) mass shell, $k \sim \pm \omega$. 
Here the superscript $r(l)$ designates right (left) moving particles.
In the former case the most singular contributions can be resummed in a usual fashion, by
introducing the self energy $\Sigma_{\omega}(k)$, and
writing the Green function in the standard form, $G_{\omega}(k) = [ \omega - k - \Sigma_{\omega}(k)]^{-1}$. The self energy can be written as (we set $v=1$ and drop the explicit $q$-dependence of $\Delta$)
\begin{align}
\Sigma_{k}(\omega)& = \frac{\Delta^2}{(\omega +k)^{1-2d}}\label{Sigma}
\\
& \times \left[C_0(d) + \sum_{n=1}^{\infty}C_n\left(d,\frac{k +\omega}{2\omega}\right)\left(\frac{\Delta^2}{(\omega +k)^{2-2d}}\right)^{n}\right] \nonumber
\end{align}
The coefficients $C_n(d,\rho)$ have singularities only at $\rho=0$ which means that the self energy is a smooth function of its arguments in the broad vicinity of the particle mass shell, $k \sim \omega$ making the resummation scheme justified. The direct calculation yields:
\bea
C_0(d,x) = \re^{-\ri\pi d} \Gamma(1-2d), ~~ 
C_{1}(d,1/2) \approx d \re^{-2\ri \pi d}\, .\label{C1}
\eea
These considerations and the fact that away from $x=0 (\omega =-k)$ $C_n(d,\rho)$ vanish at $d \rightarrow 0$ allows one to  approximate the Green function as 
\begin{equation}
G_{\omega}(k) =\left[\omega -k - \frac{\Delta^2(q)  a^{2d} }{(\omega +k)^{1-2d}}(1 - \ri \pi d )\right]^{-1} \, .
\label{G-result1}
\end{equation}
It follows from equation~\eqref{G-result1} that the quasi-particle dispersion relation, $E(k)$  as determined by 
the relation $E(k) = k + \mathrm{Re}\Sigma(E(k),k) $ differs from the mean field BSC result.
In particular, the spectral gap  identical to the Kondo scale in the magnetic impurity problem. The latter one is given by
\begin{equation}\label{gap}
T_{K} = \Delta(q) \left[\Delta(q) a\right]^{d/(1-d)} 
\end{equation}
The gap magnitude for a given $q$ 
is suppressed with respect to its mean field value $\Delta(q)$.  
In addition, the spectral line acquires a finite width, proportional to the spectrum renormalization $\Gamma(k) \approx \pi d (E(k) - k)$ at $d \ll 1$. 

\begin{figure}[h]
\begin{center}
\includegraphics[width=1.0\columnwidth]{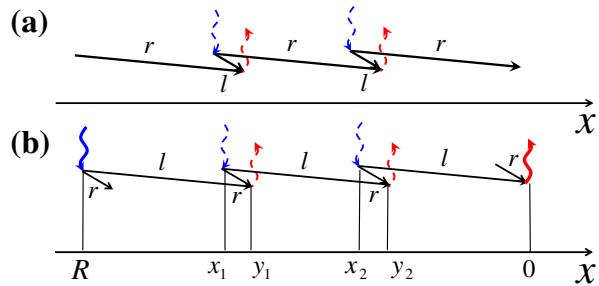}
\caption{(Color online) Propagation of the right-moving particle. (a) $k \sim  +\omega$ and (b) $k \sim  -\omega$.  
Solid straight arrowed (black) lines labeled with letters $r$ ($l$) 
designates propagators of the right-moving particle (left-moving hole).
Pairs of outgoing (red) and incoming (blue) dashed wavy lines designates the insertion of $\Delta \re^{\ri \phi}$ and $\Delta \re^{-\ri \phi}$ phase factors with close arguments.  In the panel (b) a pair of distant solid  wavy lines depicts the two uncompensated phase factors.}
\label{fig:scatt}
\end{center}
\end{figure}

We now turn to a discussion of the hole-like region $k \sim - \omega$. 
The self energy is not a useful quantity in this case as it diverges at $k = - \omega$ and becomes more singular with increasing order of the perturbation theory. 
We therefore sum the most singular contributions to the Green function. For the ``amputated'' Green function, 
$\bar{G}^{r}_{\omega}(k) = [ G_{\omega}(k) -  G_{0,\omega}(k) ]  [G^{r}_{0,\omega}(k)]^{-2}  $, we obtain the  expansion in even powers of $\Delta$ similar to 
 (\ref{Sigma}). The coefficients of this expansion are denoted as $\tilde C_n$. %
The coefficients with $n \geq 1$ are non-analytic at $k = -\omega$. To find the momentum dependence of the Green function at $k \sim -\omega$ we have calculated the whole set of constants $\tilde C_n(d,0)$ and summed over the corresponding series. 
%
%
As a result of interaction with the field created by the fluctuating order parameter the right moving particle is  scattered off as a left moving hole, see Fig.~\ref{fig:scatt}. 
When the order parameter is fixed these processes  give rise to a usual BCS-like quasi-particle spectrum.  
In our case the order parameter has different phases at different collision events. Our goal is to study the effect of these fluctuations to the leading order in the parameter $d$. For $\omega \sim  k$ the right-moving particle propagates along much longer distance in between the consecutive scatterings than the left moving hole. This is shown in Fig.~\ref{fig:scatt}(a). As a result the space arguments in the phase factors attached to the propagator of the hole merge. 
 For that reason the integration over the short distances of propagation of the hole leads to effective fusion of its propagator. This explains why the Green function is determined by the self energy to the second order in $\Delta$, i.e. the result of Eq.~\eqref{G-result1}. Indeed, corrections to Eq.~\eqref{G-result1} results from the interaction between distant dipoles and are small in parameter $d \sim T$.  

 The situation at  $\omega \sim  -k$ is different, see Fig.~\ref{fig:scatt}(b). In this case the left-moving hole propagates much longer distances than the right-moving particle. As a  result the arguments of the two outermost phase factors are separated by large distance which substantially modifies the infrared behavior of the propagator. The two uncompensated phase factors lead to a power law dependence in Eq.~\eqref{G-result2} absent in Eq.~\eqref{G-result1}.
We now derive the result of Eq.~\eqref{G-result2} analytically. To illustrate  the calculation we consider  the order $\Delta^{6}$, (see Fig.~\ref{fig:scatt}(b)): 
\begin{align}\label{amputated}
\delta \bar{G}^{r}_{\omega} &(R) =  
(-\ri)^{6} \Delta^{6} G^{l}_{0,-\omega}(y_{1} - R  )  G^{r}_{0,\omega}(y_{1}-x_{1}) \notag \\
 \times &  G^{l}_{0,-\omega}( y_{2} -x_{1} ) G^{r}_{0,\omega}(y_{2}-x_{2}) G^{l}_{0,-\omega}(-x_{2}) \notag \\
\times & \langle \re^{-\ri \phi(R) } \re^{ \ri \phi(y_{1})}   \re^{ - \ri \phi (x_{1}) } \re^{  \ri \phi (y_{2}) }  \re^{ \ri \phi(x_{2}) }   \re^{ - \ri \phi (0)} \rangle \, . 
\end{align}
Here the bare retarded Green functions $ \pm \ri G^{r,l}_{0,\pm\omega}(x) = \theta(x) \re^{\ri \omega x}$. The infrared singularity in the Fourier transformation of Eq.~\eqref{amputated} at $k \sim - \omega$ accumulates at large distances, $R \propto - |\omega + k|^{-1}$. In terms of rescaled variables, $x_{1,2} = - R \xi_{1,2}$,  $y_{1,2}  - x_{1,2} = -R \eta_{1,2}$ it means that the leading contribution comes from $\eta_{1,2} \ll 1$. Integrating over negative $R$'s we obtain
\begin{align}\label{amputated1}
\delta^{(6)}  & \bar{G}^{r}_{\omega} (k) \approx  
\frac{\Delta^{6}  \Gamma(5 - 6 d) }{ (2 \omega)^{5 - 6d} }  
\int_{0}^{1} d x_{1} \int_{0}^{x_{1}} d x_{2} \notag \\
&\times \int_{0}^{\infty} d\eta_{1}d\eta_{2} 
\frac{ \eta_{1}^{-2d} \eta_{2}^{-2d} }{ ( \rho + \eta_{1} + \eta_{2} )^{5 - 6d}   } \, .
\end{align}
As the integrals over $\eta$ variables are convergent allowing us to replace the upper integration limit by infinity. The remaining integration in Eq.~\eqref{amputated1} yields 
\begin{align}\label{amputated3}
\delta^{(6)} & \bar{G}^{r}_{\omega} (k) \approx  
\frac{\Delta^{6} \Gamma^{2}(1-2d) \Gamma(3-2d)}{ 2! (2 \omega)^{5 - 6d} {\rho}^{3-2d}}  \, .
\end{align}
The arguments leading to Eq.~\eqref{amputated3} can be generalized to obtain the most singular contribution to arbitrary order  
$\Delta^{2n}$ as follows
\begin{align}\label{amputated4}
\delta^{(2n)} & \bar{G}^{r}_{\omega} (k) \approx  
\frac{\Delta^{2n} \Gamma^{n-1}(1-2d) \Gamma(n-2d)}{ (n-1)! (2 \omega)^{2n(1 - d)-1} {\rho}^{n-2d}}  \, .
\end{align}
Summing all leading singularities  Eq.~\eqref{amputated4} we obtain 
\begin{align} \label{G-result2}
G_{\omega}(k) \simeq \frac{ 1 }{ 2 \omega }
\left[ 1  +  \frac{ \Delta^2(q)\Gamma(1-2d)\re^{-\ri\pi d} } 
{  2 \omega \left(  \omega +k -
                                                 \frac{ \Delta^2(q) \Gamma(1-2d) \re^{-\pi \ri d}   }{ (2\omega)^{1 - 2d} } 
                 \right)^{1 -2d}   
}    
\right] \, .
\end{align}
At $d \rightarrow 0$ \eqref{G-result2} reproduces the BCS Green function. \eqref{G-result1},\eqref{G-result2} are the main results of the paper.
\begin{figure}
\centering
\includegraphics[width=1.0\columnwidth]{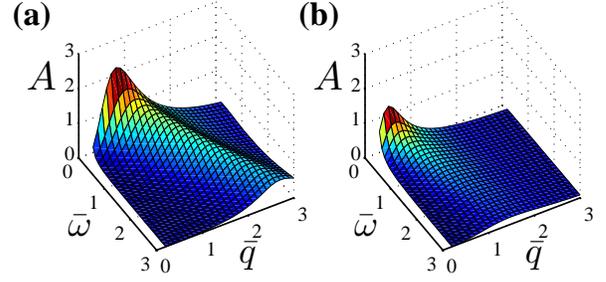}
\caption{ (Color online) 
The spectral function at $k=0$ as a function of dimensionless parameters $\bar{q}= \Delta (qa\xi/v)$ and $\bar{\omega} = (\omega\xi/v)$ as found form Eq.~\eqref{opposite2} for a) d= 0.125 and  b) d= 0.5. } \label{fig:A}
\end{figure}
%
%
%
%
%
These results remain qualitatively correct above the BKT transition, provided that the inverse correlation length $\xi^{-1}(T)$ exceeds the Kondo scale $T_{K}$ given in Eq.~\eqref{gap}. Since the BKT correlation length is exponentially large
$
\xi^{-1}(T) \sim \Delta_0\exp \left[-C(T/T_{\mathrm{BKT}} -1)^{-1/2} \right], 
$ 
 there is a range of temperatures and $q$ where this condition is fulfilled. 
In the opposite limit $\xi^{-1}  \gg \Delta(q)$ one can  use   
Eq.~\eqref{2point}) to calculate the leading order contribution to the self energy,
%
\begin{figure}
\includegraphics[width=0.7\columnwidth]{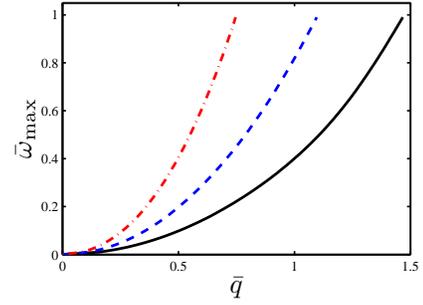}
\caption{(Color on line) The dependence of the frequency   $\bar{\omega}_{\mathrm{max}}$ maximizing the spectral function ~\eqref{opposite2} on the  dimensionless momentum $\bar{q}$ for
$d=0.5$ solid line (black), $d=0.25$ dashed line (blue), and $d=0.125$ dashed dotted line (red).} \label{max}
\end{figure}
The following formula provides an interpolation between $T > T_{\mathrm{BKT}}$ and $T <T_{\mathrm{BKT}}$ region: 
%
\begin{align}\label{opposite2}
\Sigma^{(2)} & (q,k,\omega) =\Delta^2(q)\xi (\xi/a)^{-2d}  (1 + (\xi(\omega +k))^2 )^{-1/2}    \\
\times & \big[ \exp\left\{-\ri\pi d  + 2d\sinh^{-1}[\xi(\omega +k)]\right\}  \notag \\
&-(1 + (\xi(\omega +k))^2 )^{-1/2} \big]  \, . \notag
\end{align}
In Fig.~\ref{fig:A} we present graphically the spectral function $A_{\omega}(k) = - (1/\pi)\mathrm{Im} G_{\omega}(k)$. The quasiparticle dispersion can be identified as the energy $\omega_{\mathrm{max}}$ where the spectral function is at its maximum. The dispersion relation obtained in this way is depicted in Fig.~\ref{max} for different values of parameter $d$. Below $T< T_{\mathrm{BKT}}$ where $\xi^{-1} =0$ one should use 
Eq.~\eqref{G-result2} instead of Eq.~(\ref{opposite2}) close to the singularity line $\omega =-k$. 

\section{conclusions}
 
In summary, we have studied  the electron Green function (and the related spectral function) in the regime of strong superconducting fluctuations. 
As is evident from Figs.~\ref{fig:A} and \ref{max}, these fluctuations affect the normal state dispersion in such a way that the maximum of the spectral function is shifted down in energy in comparison with its mean field value. Naturally, the effect is  more pronounced for larger temperatures. This indeed may create the impression that the system develops a ``Fermi arc''.  We hope that the advances in the experimental techniques will allow for a detailed comparison with the present theory.  
We argue that the quasi-classical approximation employed in the previous publications \cite{Millis},\cite{Dorsey},\cite{Berg} cannot provide a quantitative information about  the spectral density along the arcs. This approximation is justified only if Green function changes on the scale smaller than the variation scale of the pairing potential. In general the former is set by the particle's mass. As the quasi-particles become massless at the node the quasi-classics is not justified when  $\Delta(k)/v  < \xi^{-1}(T)$. 
More specifically the inverse square root singularity reported in Ref.~\cite{Berg} is  an artifact of the quasi-classical approximation and is in fact smeared.  The typical scale of the spectral function is temperature dependent and can be estimated as $ \sim d(\xi/v)\Delta^{2}(v/\xi) $
close to the node.

Our approach is easely generalizable for other problems where quasiparticles coexist with critical or almost critical collective excitations such as magnetic fluctuations at the onset of antiferromagnetic order.

\begin{acknowledgments}

We are grateful to A. Chubukov and Z. Tesanovic for encouraging 
discussions and interest to the work. 
A. M. T. was supported by the Center for Emerging Superconductivity funded by the U.S. Department of Energy, Office of Science. 
M. Khodas acknowledges support from BNL LDRD grant  08-002.

\end{acknowledgments}

\end{document}